\documentclass[]{aipproc}
\layoutstyle{6x9}
\begin{document}

\section*{Preface}

Planetary and brown dwarf companions to evolved stars have only recently been 
discovered. The aim of this conference was to discuss observational results 
and techniques for their detection, explore theoretical predictions for the
formation and the fate of substellar objects orbiting evolving stars, and assess
their potential impact on the evolution of the host stars. It was central to the
conference to explore the importance of new and upcoming space missions like
Kepler, GAIA and PLATO for this emerging field. 

The meeting, held at the Harmonies\"ale of the city of Bamberg, Germany from
August 11\,--\,14, 2010, was organised by the Dr.\ Remeis-Sternwarte and the
University of Erlangen-N\"urnberg. The meeting attracted 68 delegates from 16
nations and 4 continents. 

The first planets outside the solar system were discovered around a pulsar.
Therefore the programme was opened by a review on the present state of research
in the field of pulsar planets. The next session focussed on the interaction of
planets with their host stars during their evolution from a theoretical point of
view, and included aspects of binary stellar evolution as well as second and
third generation planet formation. 

In the third session the results from surveys of planets around giant/massive
stars were presented. The inventory of exoplanets around G-K giants as well as a
very hot planet transiting a rapidly rotating A star were highlighted. Stellar
oscillations in the planet hosting giant stars were discussed as a major
obstacle for the discovery of planets.

The conference programme retraced the sequence of stellar evolutionary 
stages where the horizontal branch (HB) follows after the red-giant stage.
The timing method to measure light travel time variations in pulsating stars
and eclipsing binaries has been very successful in finding substellar companions
to HB stars. Several talks reported on such discoveries, including the detection
of a planetary companion around a metal-poor star with extragalactic origin. 

The formation and existence of planetary systems around white dwarf stars were 
the subjects of the last two sessions. The first part was devoted to debris and
gaseous disks around white dwarfs, while the second part focussed on planets and
brown dwarfs orbiting white dwarfs, which included evidence for two planets
orbiting the post-CE binary NN Ser. 

The scientific programme ended with a lively round table discussion and
concluded with the decision to hold a second meeting on this subject in 2012.
The meeting finished with a reception at the Dr. Remeis-Sternwarte.  

The philosophy of the meeting was to provide an equal platform for all delegates,
and especially to foster the full participation of young scientists. We are
confident that we have achieved this goal. Invited reviews were scheduled at the
beginning of the sessions to set the stage for the contributed talks. Ample time
for discussion was provided following each talk, and extra discussion sessions
concluded the sections individually as well as the entire meeting.

We thank the Lord Mayor of Bamberg for his patronage and for the reception of the
participants in the beautiful surroundings of the ``Geyers\-w\"orth'' castle by the
Mayor Werner Hipelius. 

We are extremely grateful to the Deutsche Forschungsgemeinschaft, the University
of Erlangen-N\"urnberg and the Erlangen Centre for Astroparticle Physics for the
generous financial support, which made possible the participation of young
researchers and facilitated the attendance of several of the key note speakers.

Last but not least we thank the Scientific Organising Committee for their valuable
help with the scientific planning and scheduling of the meeting.\\

{\it Scientific Organising Committee:}\\
Noam Soker \& Uli Heber (co-chairs), Matt Burleigh (UK), Stephan Geier (Germany),
Paul Groot (The Netherlands), Artie Hatzes (Germany), Michael Jura (USA),
Mukremin Kilic (USA), Orsola De Marco (Australia), Pierre Maxted (UK),
Gijs Nelemans (The Netherlands), Sonja Schuh (Germany), Roberto Silvotti (Italy).\\
 
{\it Local Organising Committee:}\\
Horst Drechsel (chair), Edith Day, Stephan Geier, Uli Heber, Andreas Irrgang,
Ingo Kreykenbohm, Thomas Kupfer, Veronika Schaffenroth, Florian Schiller,
Fritz-Walter Schwarm.\\

{\it Patronage:}\\
Andreas Starke, Lord Mayor of Bamberg.\\

{\it Sponsors:}\\
Deutsche Forschungsgemeinschaft (DFG),
Friedrich-Alexander-Universit\"at Erlangen-N\"urnberg,
Erlangen-Centre for Astroparticle Physics.

\end{document}